# Interpretation of Quantum Theory

# An overview

Dimitris K. Lazarou

University of Athens, Greece
Physics Department
Nuclear and Elementary Particle Physics section

May 2009

**Abstract**

Quantum Mechanics, almost 80 years after its arrival, is a well established and experimentally not falsified theory. It has predicted and explained a whole series of natural phenomena of a very delicate nature. But its interpretation has not gained universal acceptance. Many scientists have considered the conceptual framework of quantum theory to be unsatisfactory. The very foundations of Quantum Mechanics is a matter that needs to be resolved in order to achieve and gain a deep physical understanding of the underlying physical procedures that constitute our world.

Email address: sagan64_gr@windowslive.com

# INTRODUCTION

The present paper is trying to give an, as clear as possible, illustration of the main categories of Quantum Theory interpretations. Philosophical notions as determinism, realism, subjectivism and empiricism are considered known, at least in principle. The same stands for physical notions as locality, contextuality, quantum collapse process and discussions about EPR paradox [1].

We will start presenting the first, historically, interpretation of QM by its founders, Copenhagen interpretation, and will continue our overview by presenting the main features, advantages, disadvantages and experimental verification or falsification, if exist, of the other interpretations (or ontologies).

Discussion here is not exhaustive of course. There are papers and books dedicated to different aspects of this foundational problem. Some of the following interpretations are interrelated and other, like the path integrals view or transactional interpretations will not be commented at all, with regard to their supporters. Special effort has been made in order to preserve the clarity of the original ideas behind the forming of the following interpretations.

**Classification of interpretations**

1.  Copenhagen interpretation
2.  Many worlds or relative state interpretation
3.  Subjectivism, consciousness' role in quantum state reduction
4.  Decoherence
5.  Objective collapse process
6.  Consistent histories approach
7.  Local hidden variables theories (LHVTs)
8.  Non-Local hidden variables theories (NLHVTs)

1.  **CONVENTIONAL (COPENHAGEN) INTERPRETATION**

Developed mainly by Niels Bohr in Copenhagen, this historically first interpretation of QM states that the measuring apparatus cannot be separate from the quantum system under investigation. Also the kind of experiment we decide to make, defines the appropriate set of eigenfunctions that will be used to construct the state of the system. Copenhagen interpretation introduced the very deep notion of complementarity, an intrinsic property of nature. This interpretation does not solve the measurement problem; how and why occurs the collapse of the wave function during the measurement process. The famous Schrödinger's cat paradox (due to linearity of the Schrödinger equation) is exactly this: if you don't open the box, you don't know the condition of the cat. But who or what decides about the cat's condition? Why the measurement apparatus behave classically? After all it is constituted of particles that are governed by QM rules. Where lies the limit between quantum and classical world? Or, there is no limit after all?

Copenhagen interpretation tells us nothing about the underlying physics of the system. It provides just the essential mathematical formalism in order to make extremely accurate predictions, to compute the probabilities of different outcomes. The state vector represents our knowledge of the system, not its physics.

## 2. MANY WORLDS OR RELATIVE STATE INTERPRETATION

This interpretation has no collapse. All possible outcomes co-exist in different branches of the 'universe'. We cannot therefore reconstruct the initial state by any means. These different branches cannot interfere or communicate in order to protect the theory itself from producing illogical situations. This theory 'resolves' the cat paradox assuming that the cat is alive in one branch and dead in the other. Also that all the observers in these branches are in the states that agree with their observation of the state of the cat: Let $|\psi_1\rangle$ be the state that cat is alive, and $|\phi_1\rangle$ the state of the observer who identifies that the cat is alive. For a dead cat, the states are $|\psi_2\rangle, |\phi_2\rangle$ in correspondence. Then we have for the whole state $|\psi\rangle$ that

$$|\psi\rangle = a|\psi_1\rangle|\phi_1\rangle + b|\psi_2\rangle|\phi_2\rangle$$

where $|a|^2 + |b|^2 = 1$.

Many worlds interpretation is suitable to those who try to describe the whole Universe with a wavefunction, assuming no external observers, and there have been serious efforts about this program.

On the contrary, this perspective has its problems. First of all is uneconomical. For many scientists the idea of an eternally splitting universe, each split for each possible outcome, is extreme. Secondly, how can anyone define the notion of probability in such a scheme (Colmogorov probabilities)? Even the creators of this interpretation hadn't been so far able to resolve it. The infinite number of universes also, can never be proved. Many worlds interpretation however preserves the deterministic character of Nature and maybe this was the main (and underlying) reason for its creation.

## 3. SUBJECTIVISM, CONSCIOUSNESS' ROLE IN QUANTUM STATE REDUCTION

This interpretation states that the ultimate and final measuring apparatus is the observer's consciousness. The cat is both alive and dead until a conscious observer opens the box and that moment the quantum superposition 'jumps' in one of the two alternatives. All that we know about ultimately about the physical Universe is the information that we perceive with our senses, our experience of the world. E.P. Wigner pointed out that the collapse of the wave function occurs that particular moment, when information enters our mind.

This is a highly philosophical position. Many philosophers agree with it. But this interpretation for (many) others is considered unsatisfactory. Our brain (although there are serious recent efforts to describe it quantum mechanically and beyond) is an instrument that evolved through geological time under the pressure of (wonderful and well established theory of) natural selection. Where lies the level of complexity of neural structure that can cause the state vector reduction? It is almost certain that consciousness and self-awareness did not exist for a long period of time on Earth. This interpretation forces us to conclude that before this event, collapse just didn't take place at all.

Another argument comes from the fact that different observers agree about the results of quantum experiments. It seems that the experimental outcome is generally independent from the presence of different conscious observers and comes in favor of the existence of an objective physical world.

## 4.     DECOHERENCE

Decoherence lies within the standard quantum formalism and introduces the idea that the so-called 'collapse of the wave function' is no longer something that actually happens. Instead, decoherence program is based at the continuous interaction of a quantum system with its environment i.e. there cannot exist any isolated quantum systems. This interaction is responsible for the transition from quantum world of superposition to the macroscopic. At the heart of decoherence program is the separation / division of the world into a number of systems and the remaining environment, and the mathematical notion of reduced density matrix.

Decoherence comes into controversy with the classical, Copenhagen, interpretation because the latter presupposes that the (macroscopic) measuring apparatuses are described by classical physics. Some thoughts have been made also about the role of the collapse process acting simultaneously with decoherence, and raising questions about the emerge of different or same preferred eigenvector basis. However we should emphasize to the fact that that decoherence mechanism has been integrated into other speculative interpretations of QM such as relative state, consistent histories and natural collapse interpretations.

Beyond that, decoherence cannot solve the problem of definite outcomes in quantum measurement. But decoherence program introduced the key idea that we should imagine a quantum system in constant interaction with its environment. In a theoretical scheme like this one, quantum entanglement is not so 'paradoxical' as it might seems. At the end of this discussion, we have to say that decoherence program is an active field of intensive research both at experimental and theoretical level.

## 5.     A RADICAL INTERPETATION: OBJECTIVE COLLAPSE PROCESS

Often referred as 'objective-R' theory [16]. 'R' stands for the quantum state reduction when a measurement is taking place. Supporters of this interpretation believe that today's QM is not here to stay and will be replaced by a theory at which R is a real process. Physicists and mathematicians who work in the field, try to balance the inconsistence of R process (discontinuous process, 'jump' from a state to another) with the unitary and simple (at least in Schrödinger's equation) time evolution of the state vector (often denoted as U-process). It is obvious that R-process is not unitary; nevertheless it coexists with U-process in standard QM formalism. Supporters of the theory believe that the apocalypse of the true procedures that happen in microscopic scale will finally resolve all the paradoxes in QM interpretation.

There has been a huge effort in this program. We will mention briefly first the two (minor) disadvantages of the family of these theories:
a) the introduction of arbitrary parameters that emerge from unknown so far physics and
b) their completely new mathematical formalism that has to be integrated to standard QM, or better, reproduces a part of it.

The two major disadvantages of the family of objective quantum reduction theories are their conflict with experiment and the violation of energy conservation.

We think that objective-R theories represent one of the most difficult to work with proposal about the interpretation of quantum theory due to mathematical and physical obstacles that have to be defeated. But if this program succeeds, the consequences will be of great importance in today's theoretical physics. Together with NLHVTs they represent a paradigm shift [15] that, we believe, will be of vital importance in the future.

## 6. CONSISTENT HISTORIES APPROACH

In this particular approach, the basic elements of reality (or 'beables') are collections of wave functions forming a subspace of Hilbert space [8]. The mathematical formalism is mainly constituted by projection operators and density matrices. The basic physical events correspond to sets of orthogonal hermitian projections which satisfy certain conditions. For a single set the physical interpretation is that a specific set corresponds to specific possible outcomes of an experimental measurement. The consistent histories formalism provides an expression for obtaining the probability of a specific result. A history is the sequence of physical events in time and one can assign probabilities to histories, a novel idea indeed.

Within the theory there exist different consistency conditions (for example Gell-Mann's and Hartle's, Griffiths' consistency conditions and other) that create consistent sets. The theory is in the need of any natural criterion that can reduce the number of sets because there are many of them in even the simplest models; there exists a large solution space. We need to know the properties and the number of the consistent sets because, by assumption, physics is described by them. Concerning the cosmological interpretation of the approach, the theory has to resolve how the projection operators corresponding to observed experimental results produce a history belonging to at least one consistent set. It is true that the formalism allows a wide range of possible views. The developers' views differ from each other. Indeed there are at least four distinct interpretations of the formalism.

There have been efforts that combine this theory with decoherence approach and also with Bohmian mechanics. It is believed that consistent histories approach illustrates the need to supplement QM with a universal selection principle in order to produce a more fundamental theory. It is very important to mention a view of a developer of the theory: 'the consistent histories formalism has taught us that there are infinitely many incompatible descriptions of the world within Quantum Mechanics'.

## 7. LOCAL HIDDEN VARIABLES THEORIES (LHVTs)

LHVTs hypothesize that there are (underlying and impossible in principle to obtain full knowledge of their values) some variables, hidden variables, that can indeed fix the precise values of all observables of a quantum system. Hidden variables are not part of the standard Hilbert space of wavefunctions. Philosophically LHVTs are fully deterministic theories in stark contrast with the intrinsically probabilistic quantum theory.

The formulation of Bell inequalities [2] (and other as CHSH inequalities, chained-Bell inequalities and Girel'son's inequalities) allowed the creation of experiments that could indeed compare the predictions of LHVTs and standard QM. A lot of experiments took place [4,5,17]. Summarizing their results, we can say that there are very strong indications against LHVTs and local realism but not a conclusive experiment due to some experimental limitations. Many people however believe that there is no need any more for a conclusive experiment and, most of all, that the plethora of experiments indicates that something must enter the scene of interpretations dramatically: the notion of non-locality.

## 8. NON - LOCAL HIDDEN VARIABLES THEORIES (NLHVTs)

In our opinion, non-locality must be an essential part of a theory beneath QM [5,17]. However its very notion constitutes a great depart from our common sense, greater than our conceptual problems with notions as superposition or collapse. We need to make clear that the notion of non-locality we discuss here (in quantum entagled states) does not transmit information between systems, does not violate causality and special relativity.

The two main NLHVTs are de Broglie - Bohm theory and Nelson's stochastic model. In the first, the hidden variable is the position of the particle which, by assumption, cannot be determined without collapse of the wavefunction. dBB model postulates the existence of a quantum potential, $Q$, of a holistic nature, which integrates non-locality and contextuality.

We begin, [3], with Scroedinger wave equation

$$i\hbar \frac{\partial \psi}{\partial t} = -\frac{\hbar^2}{2m}\nabla^2\psi + V(x)\psi,$$

where $\psi$ can be written as

$$\psi = Re^{iS/\hbar},$$

where $R, S$ are real. We substitute $\psi$ into Schroedinger equation, do the math and separating real from imaginary part obtaining the following time-evolution equations for $R, S$:

$$\frac{\partial R}{\partial t} = -\frac{1}{2m}\left(R\nabla^2 S + 2R\nabla S\right),$$

$$\frac{\partial S}{\partial t} = -\left(\frac{(\nabla S)^2}{2m} + V(x) - \frac{\hbar^2}{2m}\frac{\nabla^2 S}{R}\right).$$

We multiply the first time-evolution equation with $R$ and define probability density function $P(x) = R^2(x)$. We obtain the continuity equation

$$\frac{\partial P}{\partial t} + \nabla\left(P\frac{\nabla S}{m}\right) = 0 \Rightarrow \frac{\partial P}{\partial t} + \nabla(Pv) = 0,$$

noting that $\frac{\nabla S}{m} = v$, velocity vector of particle ($S$ is a solution of Hamilton-Jacobi equation).

Using our definition for $P(x)$, we rewrite the second time-evolution equation as

$$\frac{\partial S}{\partial t} + \frac{(\nabla S)^2}{2m} + V(x) - \frac{\hbar^2}{4m}\left(\frac{\nabla^2 P}{P} - \frac{1}{2}\frac{(\nabla P)^2}{P^2}\right) = 0.$$

We arrived at the quantum potential $Q$ we discussed above:

$$\boxed{Q(x) = -\frac{\hbar^2}{4m}\left(\frac{\nabla^2 P}{P} - \frac{1}{2}\frac{(\nabla P)^2}{P^2}\right) = -\frac{\hbar^2}{2m}\frac{\nabla^2 R}{R}.}$$

As we mentioned before, particles possess a real position $x$ and velocity $v$. Here, $x$ and $v$ are precisely definable and continuously varying variables. The character of the wavefunction is not just mathematical but this 'matter wave' exists and guides the particle's motion. The field $V(x) + Q(x)$ exerts a force on the particle in a way that is analogous to the way in which an electromagnetic field exerts a force on a charged particle.

This model has been investigated thoroughly in many different physical situations (scattering, many – body states, transitions between stationary states, etc.). dBB model reproduces all the results of ordinary QM but has not gained universal acceptance for the following reasons:
   a) the quantum potential has no physical basis
   b) it is a non-local theory
   c) there emerge serious problems integrating special relativity

d) physical properties of the particles (such as electric dipole moment or gravitational mass) come in direct contradiction with those we measure if we apply dBB model in a single electron of a hydrogen atom for example.

We will comment in brief the above statements:
a) We are already in a physics domain that we perceive it as a 'black box': $|initial\rangle$ to $|final\rangle$ state. In the middle there is a black box, we do not know the delicate physical procedures that take place and may give rise to Q. In our opinion, QM is a very ingenious way to handle these difficulties. After all we consider an area of $10^{-18}$m at most, beneath this (not to mention Planck scale) the physical procedures might be of a completely different nature. There are proposals like this [11], considering a deterministic theory at Planck scale and a loss of information as we come to larger scales, giving birth to hidden variable concepts.
b) Non-locality should be a fundamental part of our future theory underlying QM. Non-locality in the form of quantum entanglement is a well established experimental fact.
c) There have been efforts about this problem, see for example [18, 19, 20]. As far as we know there is no complete solution.
d) This is the main problem of dBB model because it contradicts (not only experimental facts but also) the principles of the theory themselves (regarding the 'real' position of the particle for example).

Nelson's stochastic model [21] represents a mentionable attempt of constructing a NLHVT. NSM is also a model where particles do possess classical trajectories. Once again positions are the hidden variables, perturbed by an underlying Brownian motion, in our opinion an underlying unknown physics that mathematically can be expressed in the form of this particular motion. One can derive from the appropriate Lagrangian

$$L(x,\dot{x}) = \frac{1}{2}m\dot{x}^2 - V(x)$$

including stochastic processes

$$dx(t) = v_{(+)}(x(t),t)dt + dw(t)$$

the appropriate quantum equation of motion, equivalent to Schrödinger's equation. $dw(t)$ is the Brownian type pertubation. Introducing a diffusion process with density $\rho(x,t)$ we can rewrite the wave function as

$$\psi = \sqrt{\rho(x,t)}e^{iS/\hbar}.$$

Once again we arrive at

$$\frac{\partial S}{\partial t} + \frac{(\nabla S)^2}{2m} + V(x) + Q(x) = 0$$

where the potential $Q(x)$ has analogous form with the previous one.

NSM can be viewed as equivalent to QM as far as we concerned about model's predictions only.

On the contrary some authors claim that dBB model and standard, Copenhagen, QM are not just different interpretations but different theories, because of some subtle differences concerning mainly the 'empty-wave' behavior of dBB model. NSM has not raised such controversies. It does not however shed light at the fundamental physical procedures that might give rise to quantum phenomena, assuming just an underlying Brownian-type disturbance. However we think that efforts like this and dBB are the first steps toward the solution of this vast conceptual problem of QM interpretation.

**EPILOGUE**

In our opinion, NLHVTs and 'objective-R' theories comprise the leader candidates for the re-interpretation and re-formulation of Quantum Theory. It is noteworthy how few physicists (comparably) rate the problem of interpretation of QM as a fundamental one. Theoretical physics community must apprehend that the program of quantizing gravity, which is considered as the main problem in today's theoretical physics, should embody the solution to QM interpretation problem, in order to envisage the deeper mechanisms of Nature. Physics should tell us not only how to predict the various experimental outcomes [10] but also what Nature *is*.